\pdfoutput=1

\documentclass[a4paper]{PoS}

\usepackage{amsmath}
\usepackage{subfigure}
\usepackage[inline,shortlabels]{enumitem}
\usepackage{lineno}
\usepackage{url}
\usepackage{float}

\title{Measurements of photon and jet production properties with ATLAS}

\ShortTitle{Photons and jets in ATLAS}

\author{\speaker{Christopher John Meyer}, on behalf of the ATLAS Collaboration\\
        University of Pennsylvania (US)\\
        E-mail: \email{chris.meyer@cern.ch}}

\abstract{
Summary of recent ATLAS measurements of jet and photon production using proton--proton ($pp$) collisions from the Large Hadron Collider.
The charged-particle multiplicity in jets, and jet charge measurements are presented using 8~TeV $pp$ collisions.
Differential measurements of jet and photon cross-sections are shown for 7, 8 and 13~TeV $pp$ collisions.
}

\FullConference{Fourth Annual Large Hadron Collider Physics\\
		13-18 June 2016\\
		Lund, Sweden}

\begin{document}
%\linenumbers

\section{Introduction}

Quantum chromodynamics (QCD) is an incredibly successful theory which describes the interaction and resulting kinematics of colored objects.
When modeling a proton--proton ($pp$) collision at the Large Hadron Collider, there are four general contributions:
\begin{enumerate*}[1)]
  \item hard-scatter,
  \item fragmentation,
  \item hadronisation, and
  \item underlying event.
\end{enumerate*}
The hard-scatter process models the scattering of the two incoming partons, and is described by perturbative QCD calculations.
Current calculations are generally accurate out to next-to-next-to-leading order (NNLO), and are also dependent on the parton distribution function (PDF) of the incoming protons.
Fragmentation takes the outgoing hard-scatter partons to the non-perturbative regime through parton emission and splitting.
Upon entering the non-perturbative regime, where quarks are confined, hadronisation evolves them to the state observed by the detector.
The underlying event, which includes color-connected beam--beam remnants and multiple-parton interactions within the colliding protons, is also described by non-perturbative models.

\begin{figure}[b]
  \centering
  \subfigure[]{\includegraphics[width=0.42\textwidth]{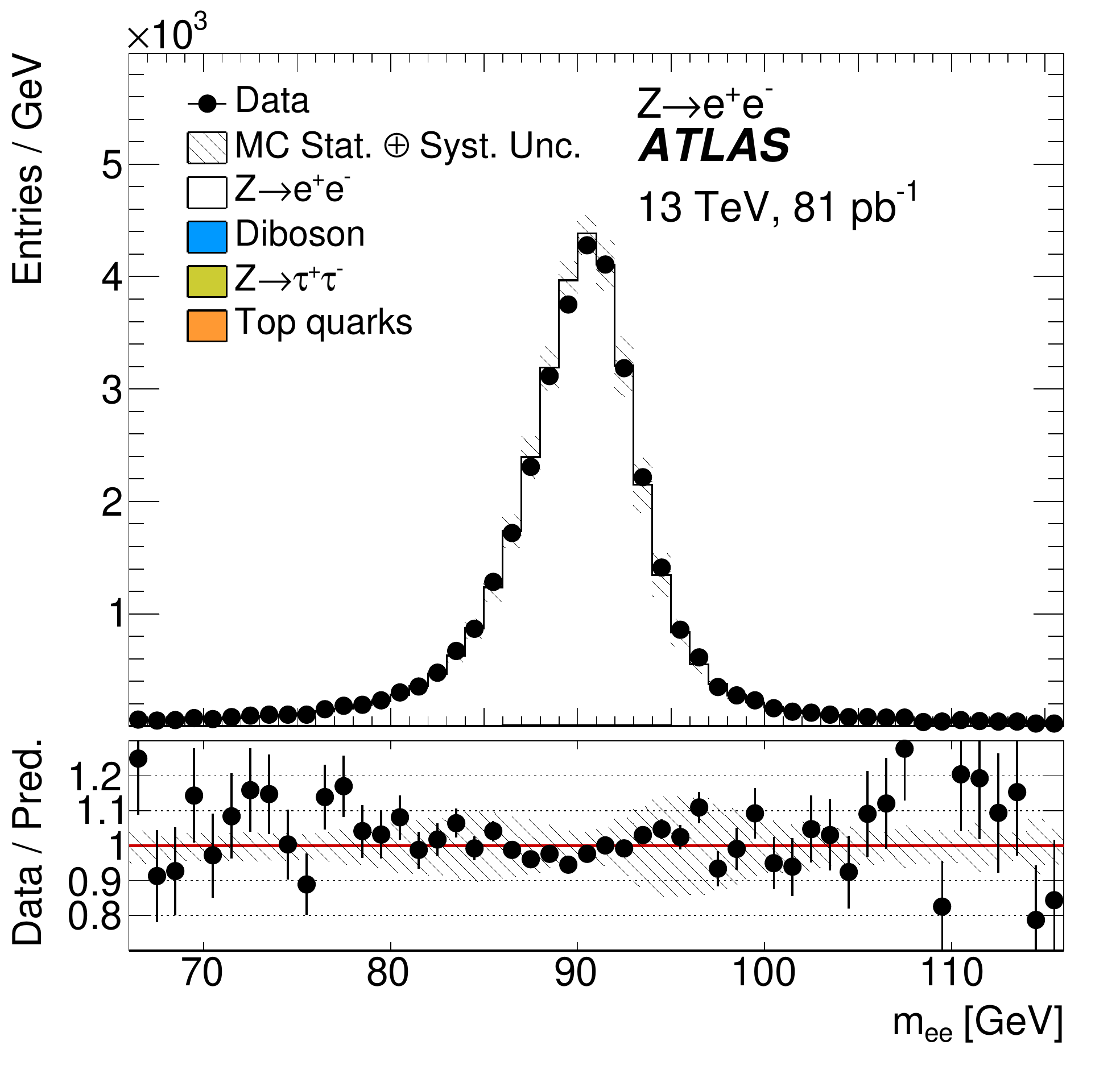}\label{fig:jetSel:a}}
  \subfigure[]{\includegraphics[width=0.56\textwidth]{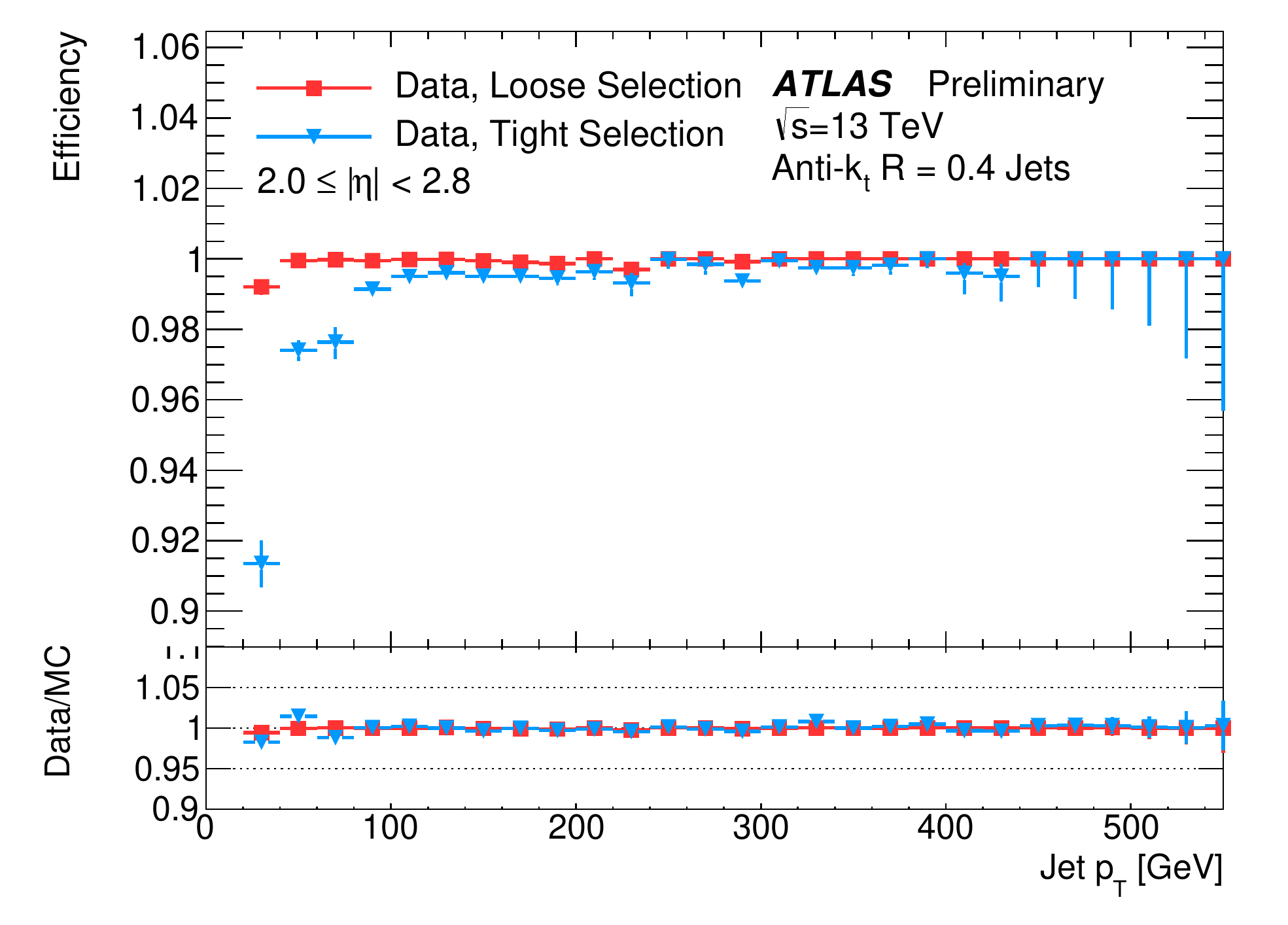}\label{fig:jetSel:b}}
  \caption{
    In (a) the calibrated dielectron mass, dominated by the $Z \rightarrow ee$ process, is shown for data and MC simulation~\cite{Aad:2016naf}.
    In (b) the absolute jet selection efficiency is shown, as well as the ratio of data and MC simulation~\cite{Aad:jetSel}.
  }
  \label{fig:jetSel}
\end{figure}

Photons provide a good measurement of the hard-scatter, constraining both the accuracy of the perturbative calculation as well as the proton PDF.
Measured in the ATLAS~\cite{Aad:2008zzm} electromagnetic calorimeter, the clusters are calibrated using a multivariate analysis trained on simulations.
An additional data-driven correction is also applied, resulting in well calibrated photons.
Figure \ref{fig:jetSel:a} shows the Z-boson decaying into two electrons, which have a similar calibration applied~\cite{Aad:2016naf}.
The data are seen to agree well with the known position of the Z-boson mass peak.

Because partons can not be measured directly, the resulting particles from the fragmentation and hadronisation simulation are clustered into \emph{particle-level jets}.
The ATLAS collaboration predominantly uses the anti-$k_t$ clustering algorithm~\cite{Cacciari:2008gp}, with matching radius $\Delta R = 0.4$.
Particle-level jets consider all charged particles with lifetime $c\tau > 30$~ps, except for muons, as inputs to the clustering algorithm.
\emph{Detector-level jets} are formed by clustering topological energy clusters from both the hadronic and electromagnetic calorimeters.
The resulting detector-level jet is further calibrated using several data-driven techniques, resulting in a well measured energy scale and resolution.
As seen in figure \ref{fig:jetSel:b}, jets with $p_\mathrm{T} > 20$~GeV have a selection efficiency $> 98\%$~\cite{Aad:jetSel}, which is well modeled by the MC simulation.

For both jet and photon measurements, detector efficiency and resolution effects are corrected (unfolded) using MC simulation, resulting in a measurement at the particle-level.
In general, the fiducial acceptance of the particle-level prediction is very similar to the event selection applied to data, which minimizes the dependence of the unfolding method on the MC simulation.
This allows data to confront PDF models, perturbative calculations, as well as the modeling of factorisation and hadronisation.

Recent ATLAS measurements of charged-particle multiplicity in jets~\cite{Aad:2016oit}, as well as jet charge~\cite{Aad:2015cua}, confront the simulation of jet properties such as flavour identification and hadronisation.
The jet charge, along with differential measurements of jet~\cite{Aad:jet13TeV} and photon~\cite{Aad:2016xcr,Aad:photon13TeV} production, provide a probe of the PDF.
Finally, differential jet and photon production are important for testing perturbative QCD calculations.

%% \begin{figure}
%% \includegraphics[width=.6\textwidth]{figures/Zee.pdf}
%% \caption{Electron cluster calibration.}
%% \label{fig:Zee}
%% \end{figure}

\section{Jet properties}

\begin{figure}[b]
  \centering
  \subfigure[]{\includegraphics[width=0.49\textwidth]{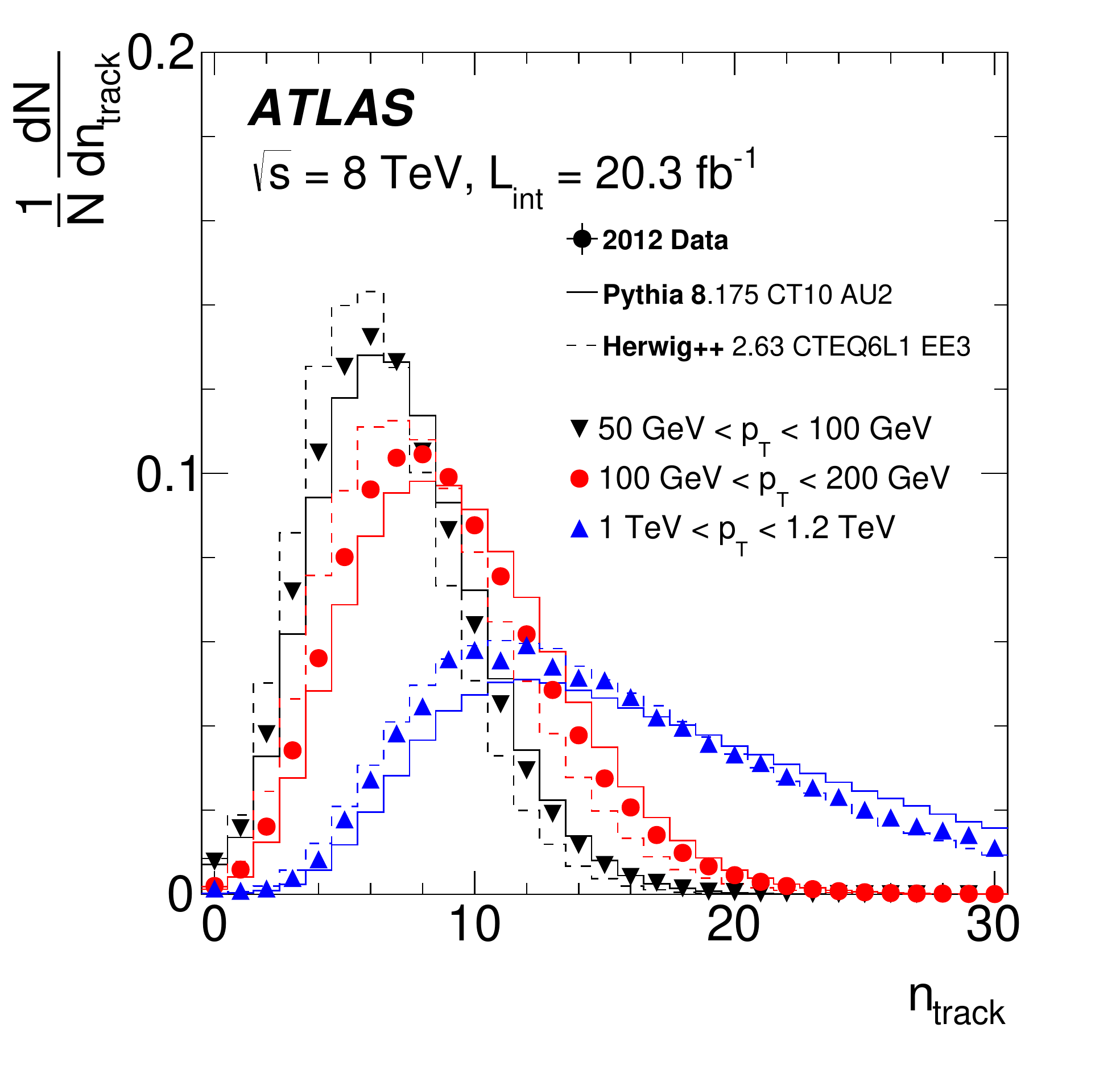}\label{fig:jetMultNtrk:a}}
  \subfigure[]{\includegraphics[width=0.49\textwidth]{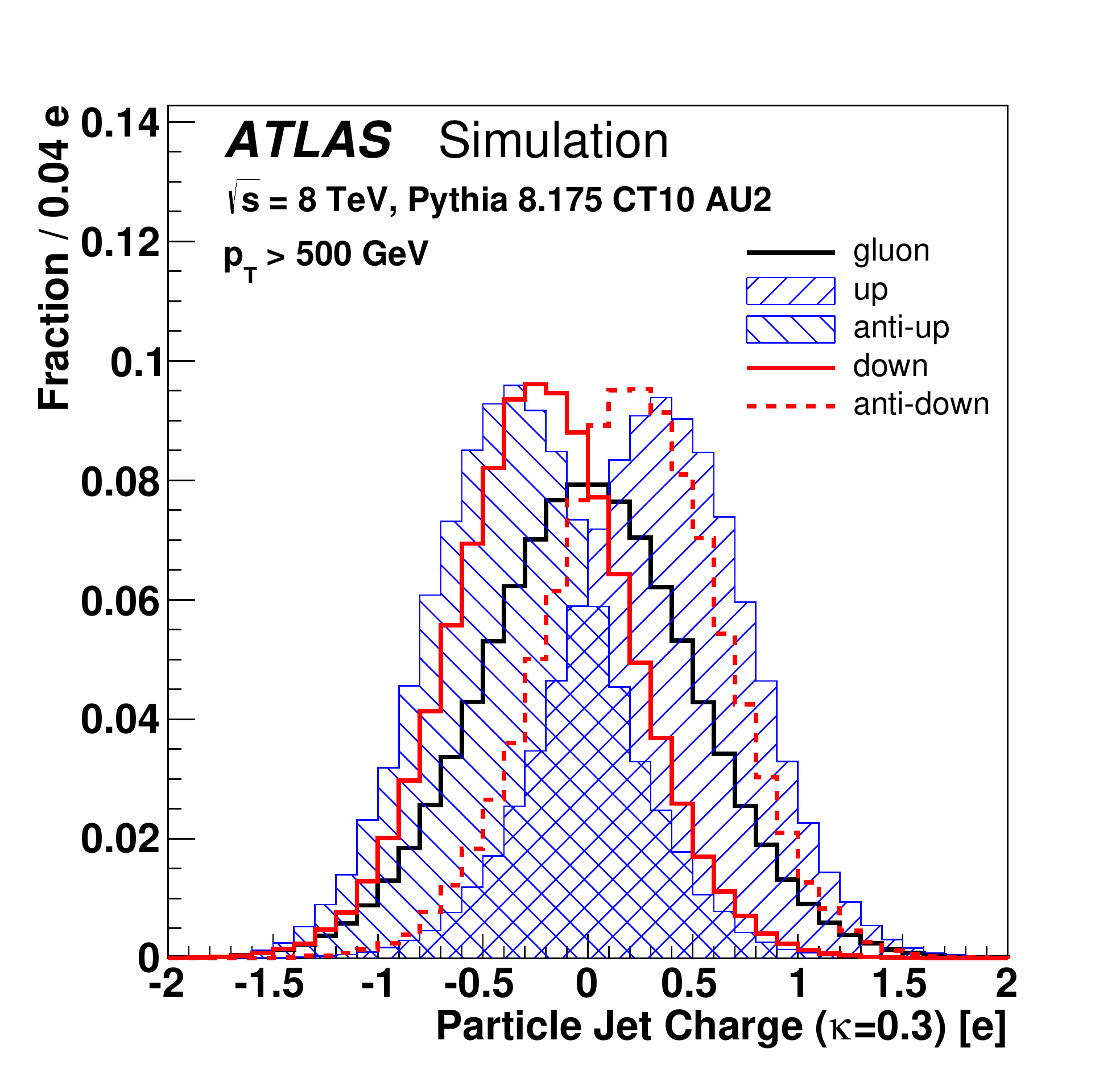}\label{fig:jetMultNtrk:b}}
  \caption{
    In (a) is the charged-particle track multiplicity for three different ranges of jet $p_\mathrm{T}$~\cite{Aad:2016oit}.
    In (b) the jet charge is shown for different flavours of parton which initiated the jet~\cite{Aad:2015cua}.
  }
  \label{fig:jetMultNtrk}
\end{figure}

Two recent ATLAS results study the properties of jets:
\begin{enumerate*}[1)]
 \item charged-particle multiplicity in jets, and
 \item jet charge.
\end{enumerate*}
Both measurements consider well-balanced dijet events produced in 8~TeV $pp$ collisions.
Individual jets are selected with absolute pseudo-rapidity $|\eta| < 2.1$ and transverse momentum $p_\mathrm{T} > 50$~GeV.
In addition, the ratio of transverse momentum of the leading jet with the sub-leading jet must be less than $p_\mathrm{T,1} / p_\mathrm{T,2} < 1.5$.
Of the two highest $p_\mathrm{T}$ jets, the jet with the smaller (larger) absolute pseudo-rapidity is classified as the \emph{central} (\emph{forward}) jet.
Due to PDF effects, the forward jets are less likely to be associated with gluons.

Within the selected jets, charged-particle tracks are considered if they fall within the inner tracking detector, $|\eta| < 2.5$, and have $p_\mathrm{T} > 500$~MeV.
An example of the charged-particle track multiplicity $\langle n_\mathrm{charged} \rangle$ is shown in figure \ref{fig:jetMultNtrk:a}, where data are compared with predictions by \textsc{Pythia8}~\cite{Sjostrand:2007gs} and \textsc{Herwig++}~\cite{Bellm:2013hwb}.
Both MC generators show differences compared to the data, over a range of jet $p_\mathrm{T}$ values.

\begin{figure}[t]
  \centering
  \subfigure[]{\includegraphics[width=0.49\textwidth]{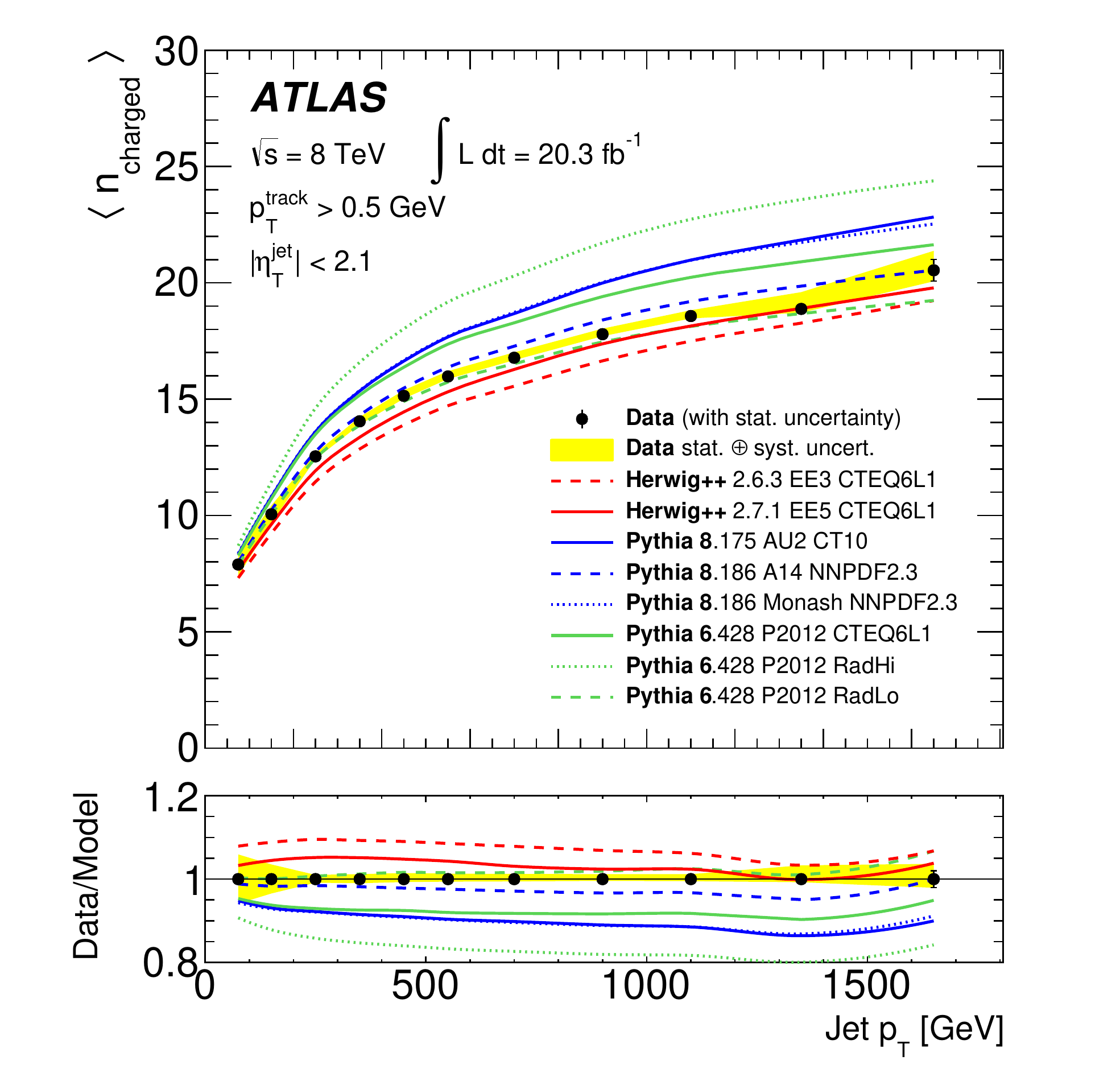}\label{fig:jetMultAvgNtrk:a}}
  \subfigure[]{\includegraphics[width=0.49\textwidth]{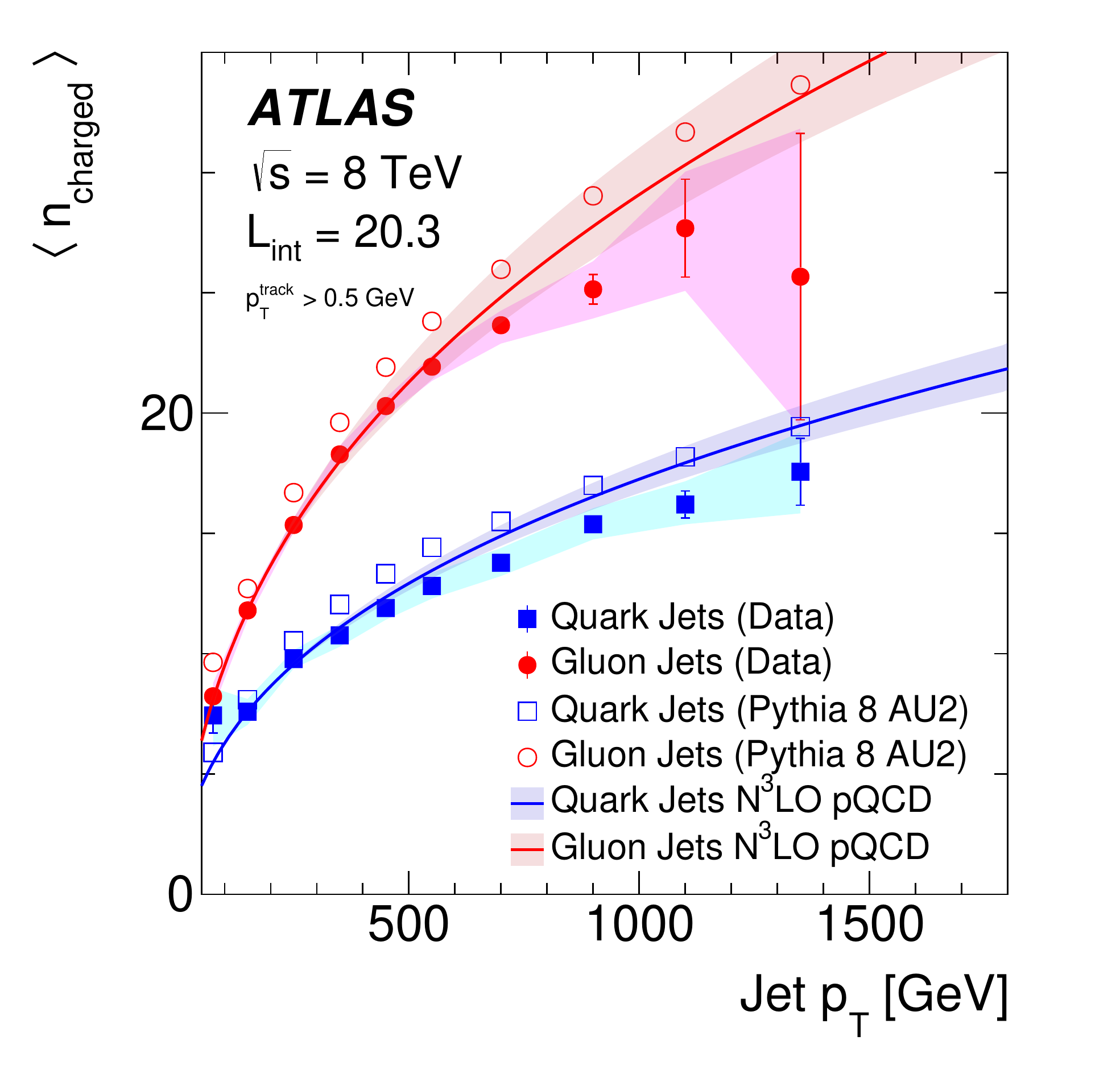}\label{fig:jetMultAvgNtrk:b}}
  \caption{
    In (a) is the average charged-particle track multiplicity as a function of jet $p_\mathrm{T}$~\cite{Aad:2016oit}.
    In (b) the average charged-particle track multiplicity is shown as a function of jet $p_\mathrm{T}$ for different flavours of parton which initiated the jet~\cite{Aad:2015cua}.
  }
  \label{fig:jetMultAvgNtrk}
\end{figure}

The average value of the charged-particle multiplicity distribution is shown as a function of jet $p_\mathrm{T}$ in figure~\ref{fig:jetMultAvgNtrk:a}.
In general, various tunes of \textsc{Pythia6}~\cite{Sjostrand:2006za} and \textsc{Pythia8} overestimate the data, while \textsc{Herwig++} underestimates the data.
Within the uncertainties, \textsc{Pythia6} using the P2012 RadLo tune shows the best description across the full jet $p_\mathrm{T}$ range.

The average charged-particle multiplicity for different jet flavours is also extracted, shown in figure~\ref{fig:jetMultAvgNtrk:b}.
This is extracted using the measured values of $\langle n_\mathrm{charged}^{c/f} \rangle$ as inputs and solving the linear equations:
\begin{align*}
\langle n_\mathrm{charged}^c \rangle &= f_q^c \langle n_\mathrm{charged}^q \rangle + f_g^c \langle n_\mathrm{charged}^g \rangle \\
\langle n_\mathrm{charged}^f \rangle &= f_q^f \langle n_\mathrm{charged}^q \rangle + f_g^f \langle n_\mathrm{charged}^g \rangle
\end{align*}
where $f_q^{c/f}$ ($f_g^{c/f}$) is the fraction of events from quark (gluon) initiated jets in the central ($c$) or forward ($f$) region, determined using MC simulation.
Two different theory predictions are considered: a leading-order (LO) prediction using \textsc{Pythia8} with the AU2 tune, and a next-to-next-to-next-to-leading-order (N$^3$LO) calculation~\cite{Capella:1999ms,Dremin:1999ji}.
While the N$^3$LO calculation better describes the data at lower values of jet $p_\mathrm{T}$, both predictions overestimate the data for larger values of jet $p_\mathrm{T}$.

Useful as a tool for flavour discrimination as well as studying hadronisation, the jet charge ($Q_J$) is defined as:
\begin{equation*}
Q_J = \frac{1}{(p_{\mathrm{T}J})^k} \sum_{i \in Tracks} q_i \times (p_{\mathrm{T},i})^\kappa
\end{equation*}
where $p_{\mathrm{T}J}$ is the transverse momentum of the jet, $q_i$ ($p_{\mathrm{T},i}$) is the charge (transverse momentum) of the track, and $\kappa$ is a regularisation parameter.
Lower values of $\kappa$ enhance the jet charge's sensitivity to low-$p_\mathrm{T}$ tracks, while in the limit $\kappa \rightarrow \infty$ only the highest-$p_\mathrm{T}$ track is considered.
An example of the jet charge distribution for different flavours of jets is shown in figure~\ref{fig:jetMultNtrk:b}.
While the mean of the different distributions clearly reflects the charge of various jet flavours, the width of the distribution lowers the expected purity when using the jet charge as a flavour discriminant.

\begin{figure}[tbp]
  \centering
  \subfigure[]{\includegraphics[width=0.49\textwidth]{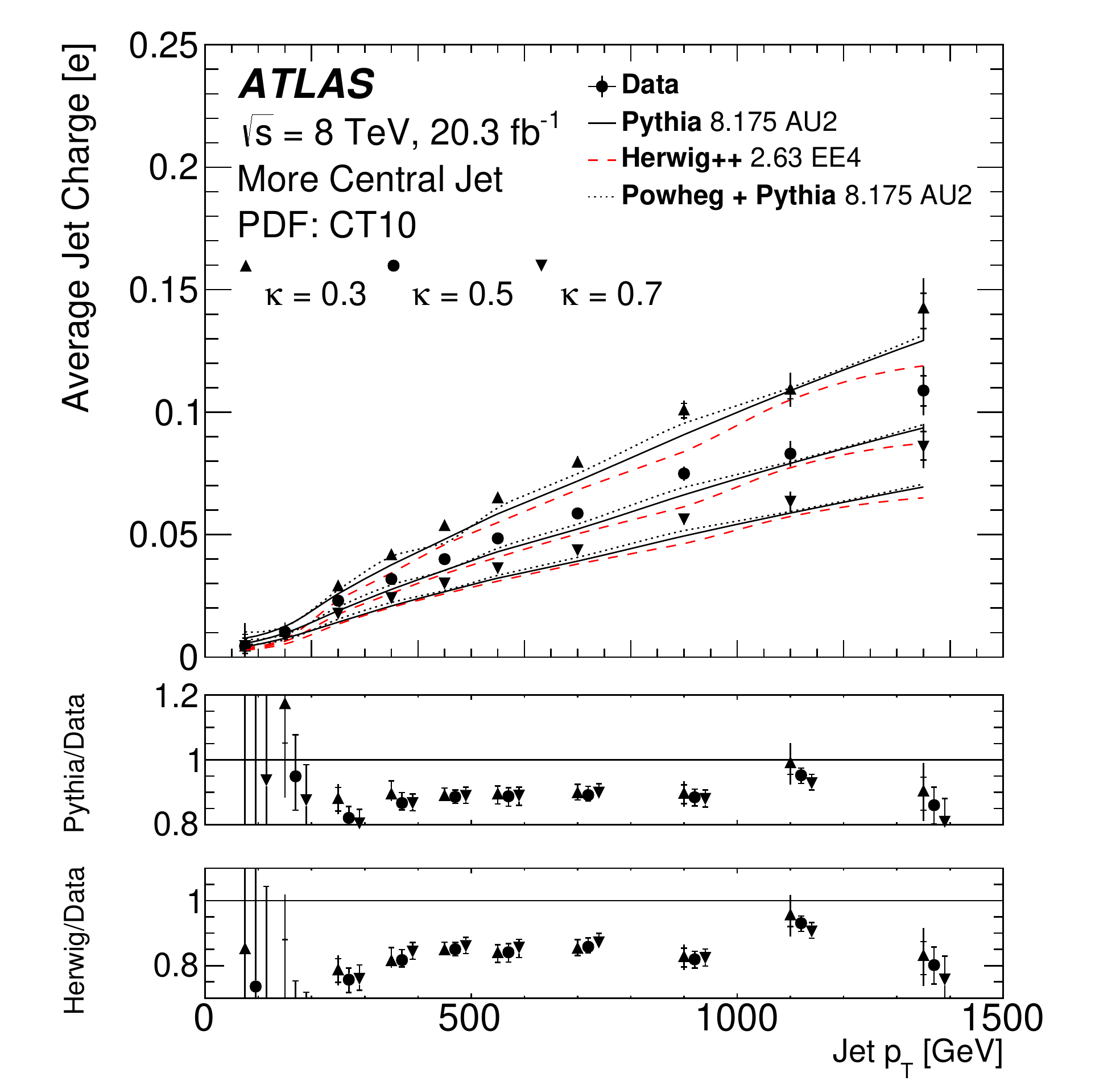}\label{fig:jetChargeFwd:a}}
  \subfigure[]{\includegraphics[width=0.49\textwidth]{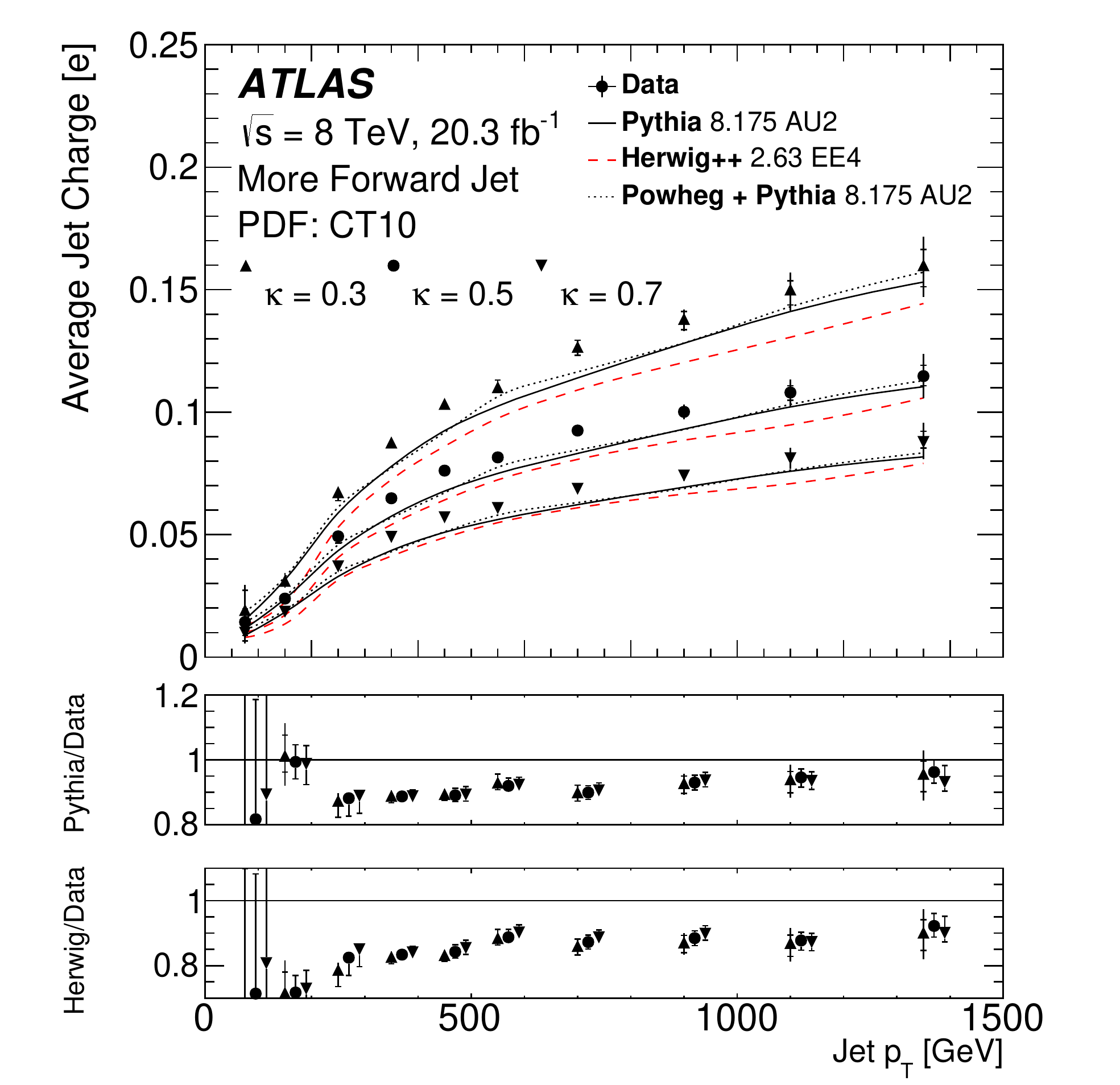}\label{fig:jetChargeFwd:b}}
  \caption{
    The average jet charge is shown as a function of jet $p_\mathrm{T}$ for (a) central jets, and (b) forward jets~\cite{Aad:2015cua}.
    For both cases, the result using three different values of the regularisation parameter $\kappa$ is shown.
  }
  \label{fig:jetChargeFwd}
\end{figure}

The average jet charge is shown in figure~\ref{fig:jetChargeFwd:a} for central jets, and in figure~\ref{fig:jetChargeFwd:b} for forward jets.
Theory predictions by \textsc{Pythia8} using the AU2 tune, and \textsc{Herwig++} using the EE3 tune, are also considered.
The \textsc{Pythia8} prediction generally follows the data better at low jet $p_\mathrm{T}$, while \textsc{Herwig++} describes the data better at high jet $p_\mathrm{T}$.

\section{Probing the hard scatter with jets}

\begin{figure}[p]
  \centering
  \includegraphics[width=0.9\textwidth]{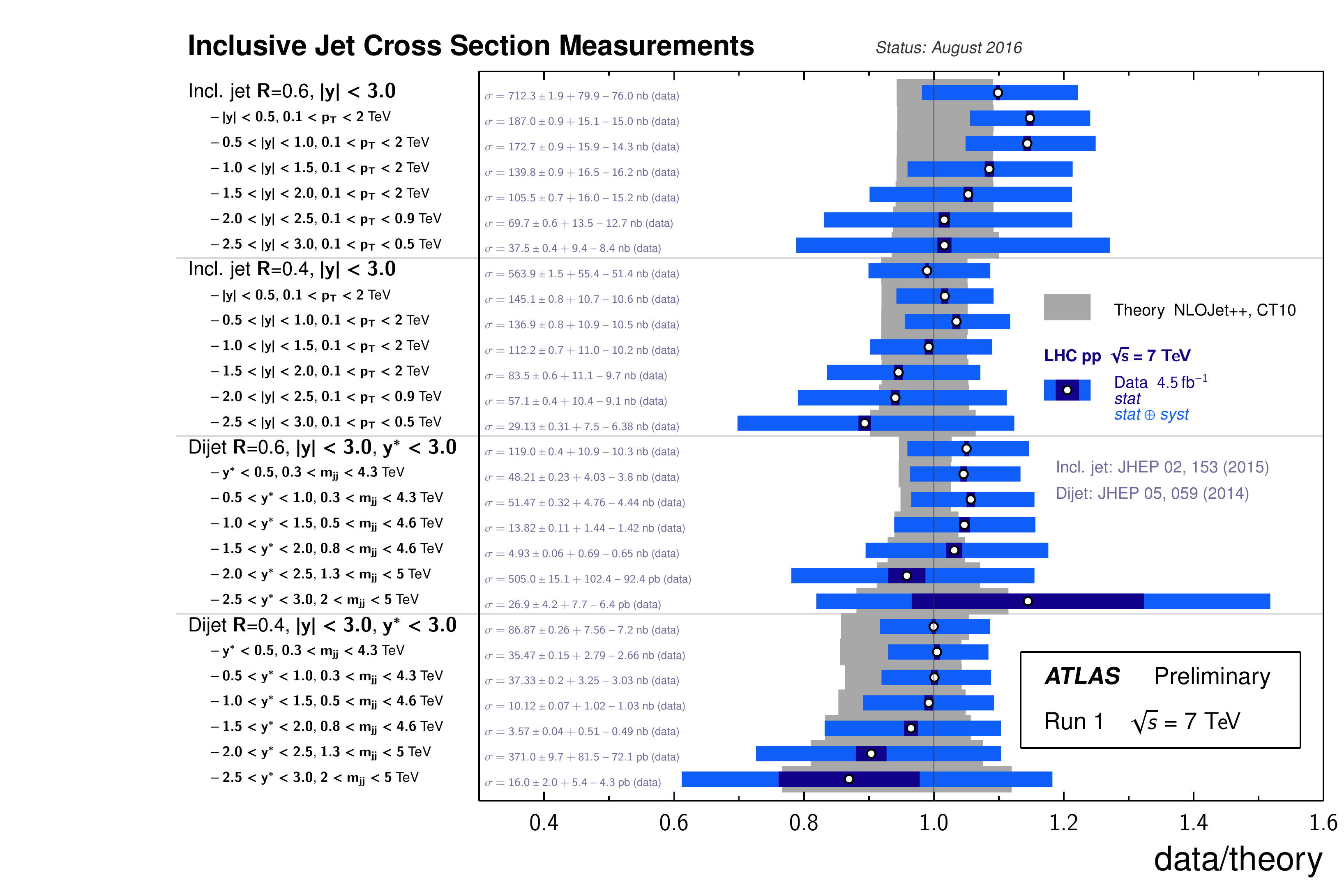}
  \caption{
    Summary of inclusive jet and dijet measurements for two values of the jet radius parameter ($R=0.4$ and $R=0.6$), compared with NLO theory predictions~\cite{Aad:jetSummary}.
    The top half of the figure shows the inclusive jet results in different ranges of jet rapidity, while the bottom half shows the dijet results in different ranges of $y^*=|y_1 - y_2|/2$.
  }
  \label{fig:jetSummary}
\end{figure}

\begin{figure}[p]
  \centering
  \subfigure[]{\includegraphics[width=0.39\textwidth]{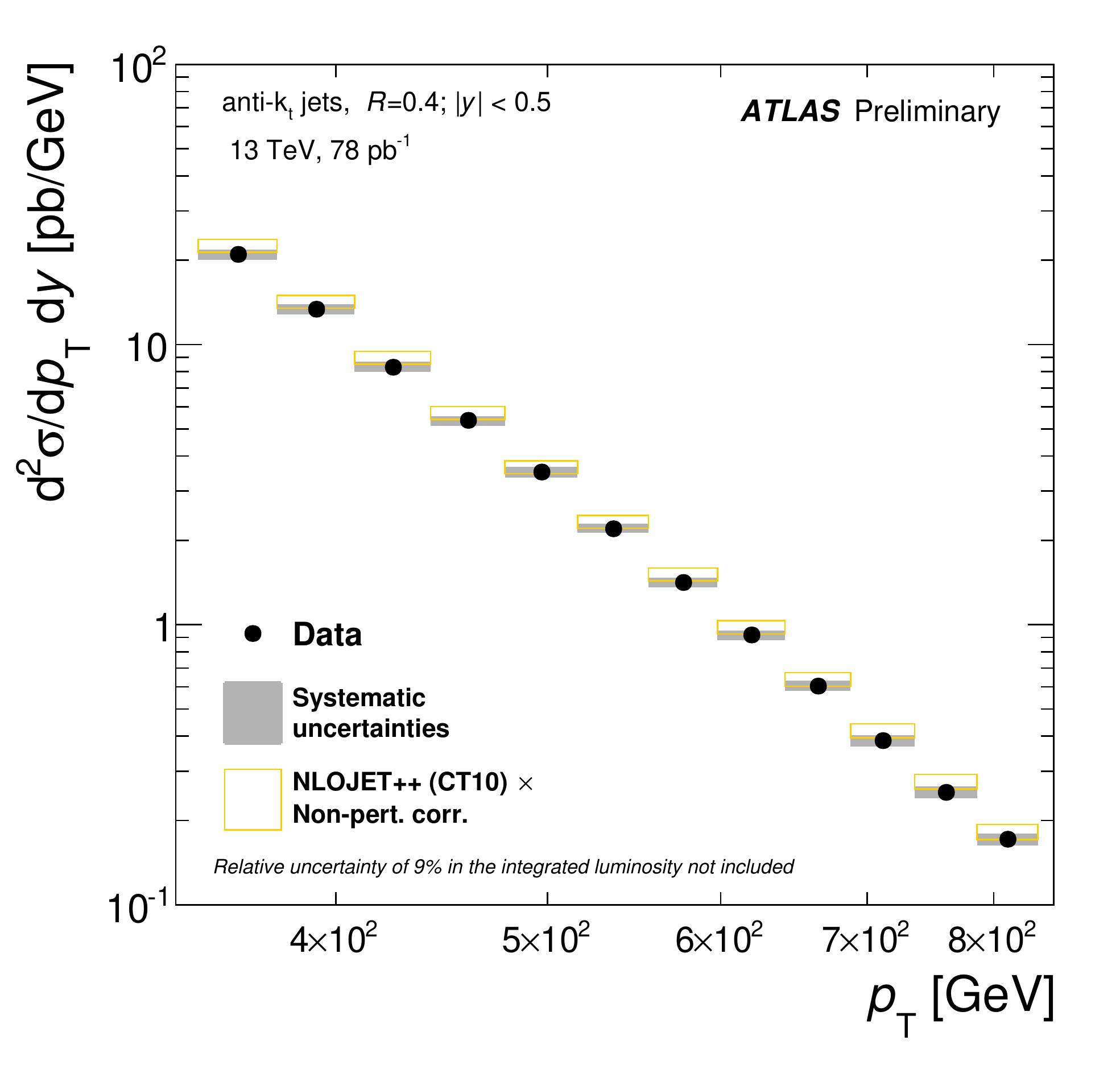}}
  \subfigure[]{\includegraphics[width=0.59\textwidth]{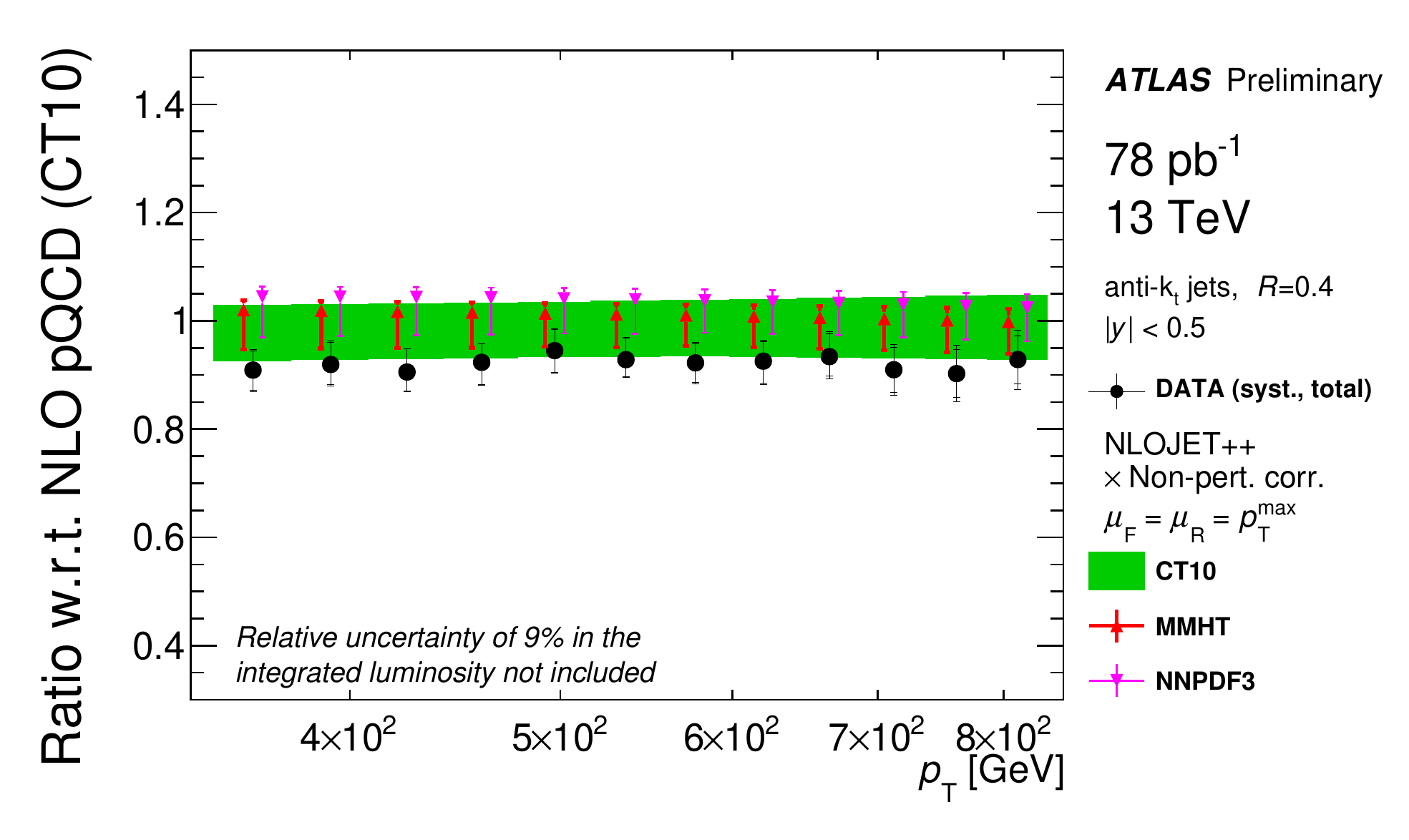}}
  \caption{
    Inclusive jet cross-sections measured with 13~TeV $pp$ collisions~\cite{Aad:jet13TeV}, compared with \textsc{NLOJet++} theory predictions using the CT10 PDF set.
  }
  \label{fig:jetAbs}
\end{figure}

Inclusive jet and dijet measurements provide a good test of perturbative QCD predictions, as well as PDF models.
A summary of measurements made using 7~TeV $pp$ collisions~\cite{Aad:jetSummary} is shown in figure~\ref{fig:jetSummary}, including NLO predictions by \textsc{NLOJet++}~\cite{Nagy:2003tz} using the CT10 PDF set~\cite{Lai:2010vv}.
An impressive agreement of the theory prediction with data is seen over a large kinematic range.
While the experimental uncertainties are currently similar to the theory uncertainties, they are expected to continue being reduced.
Recent updates to various PDF sets already reduce their component of the theory uncertainty, which is now approaching the same magnitude as the renormalisation and factorisation scale uncertainty in some kinematic regions.
This coincides well with the expected release of NNLO predictions in the near future, which will further reduce the theory uncertainty.

A recent differential measurement of inclusive jet production using 13~TeV $pp$ collisions is shown in figure~\ref{fig:jetAbs}.
In this preliminary measurement of low-$p_\mathrm{T}$ central jets ($|y| < 0.5$), the theory describes the data well within the assigned uncertainties.
The CT10, MMHT~\cite{Harland-Lang:2014zoa}, and NNPDF3~\cite{Ball:2014uwa} PDF sets are considered for the \textsc{NLOJet++} calculation.

\section{Probing the hard scatter with photons}

\begin{figure}[tb]
  \centering
  \subfigure[]{\includegraphics[width=0.55\textwidth, trim = 0cm -0.8cm 0cm 0cm]{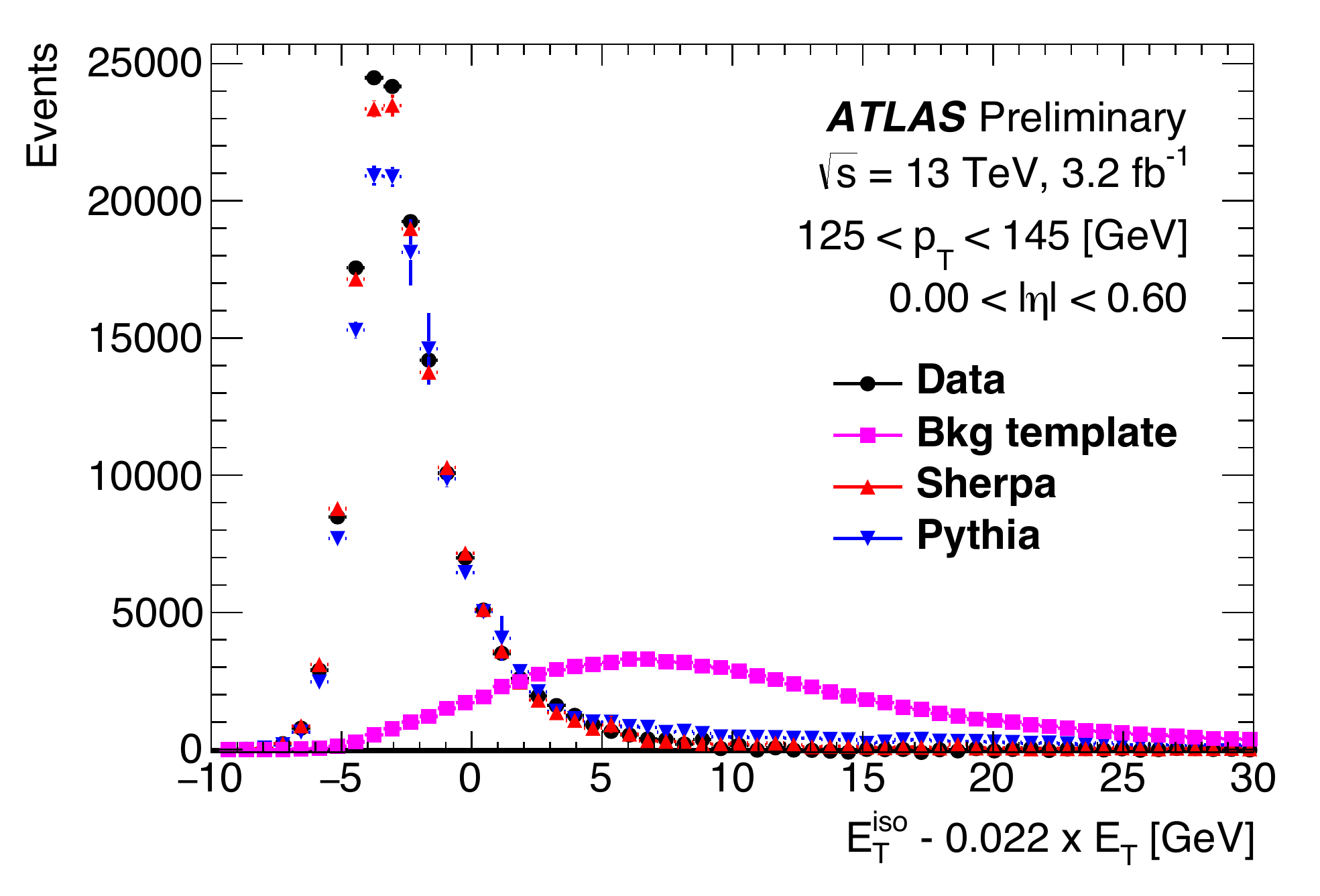}\label{fig:photonIso:a}}
  \subfigure[]{\includegraphics[width=0.42\textwidth]{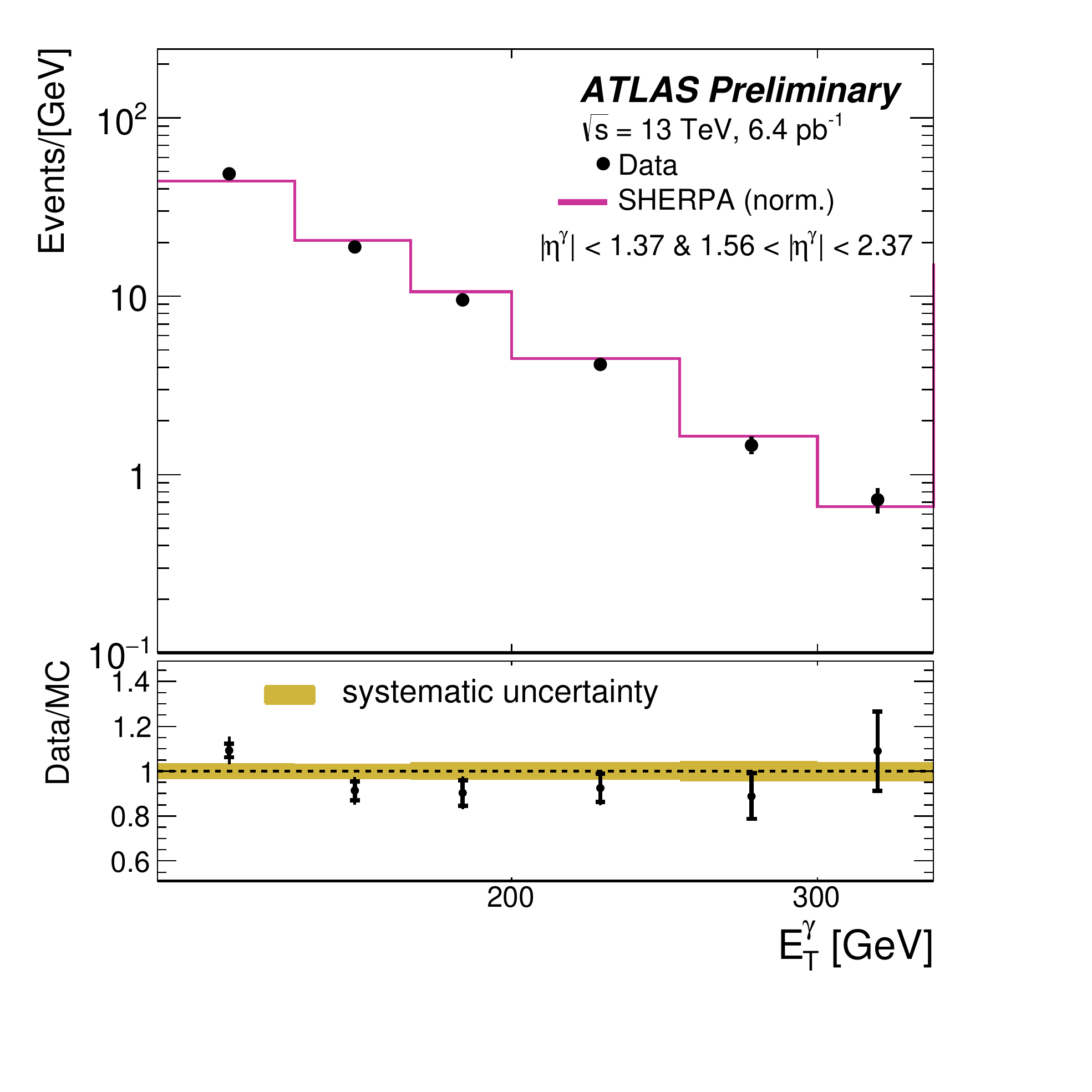}\label{fig:photonIso:b}}
  \caption{
    In (a) the photon isolation profile from 13~TeV $pp$ collisions is shown~\cite{Aad:photonIso}, with the background template in purple, the background subtracted data in black, and signal photon isolation profiles from \textsc{Pythia} and \textsc{Sherpa}.
    In (b) is a preliminary energy distribution of isolated photons~\cite{Aad:photon13TeV}, again using 13~TeV $pp$ collisions, compared with the shape prediction made by \textsc{Sherpa}.
  }
  \label{fig:photonIso}
\end{figure}

Inclusive photon production provides a similar test of perturbative QCD and PDF models.
Both photons produced in the matrix element (direct production) and during the fragmentation process are considered.
To reduce the background resulting from photons produced during hadronisation, an isolation requirement is applied.
An example of the isolation profile of photons produced in 13~TeV $pp$ collisions is shown in figure~\ref{fig:photonIso:a}, where the isolation of the signal photons is shown after subtracting the background template~\cite{Aad:photonIso}.

A preliminary differential measurement of photon transverse energy unfolded to the particle-level is shown in figure~\ref{fig:photonIso:b}.
Photons with absolute pseudo-rapidity $|\eta| < 2.37$ are considered, excluding the crack in the detector between $1.37 < |\eta| < 1.56$ where photons are reconstructed poorly.
Within the uncertainties, the LO \textsc{Sherpa}~\cite{Gleisberg:2008ta} prediction describes the shape of the data.

\begin{figure}[ht]
  \centering
  \includegraphics[width=0.9\textwidth]{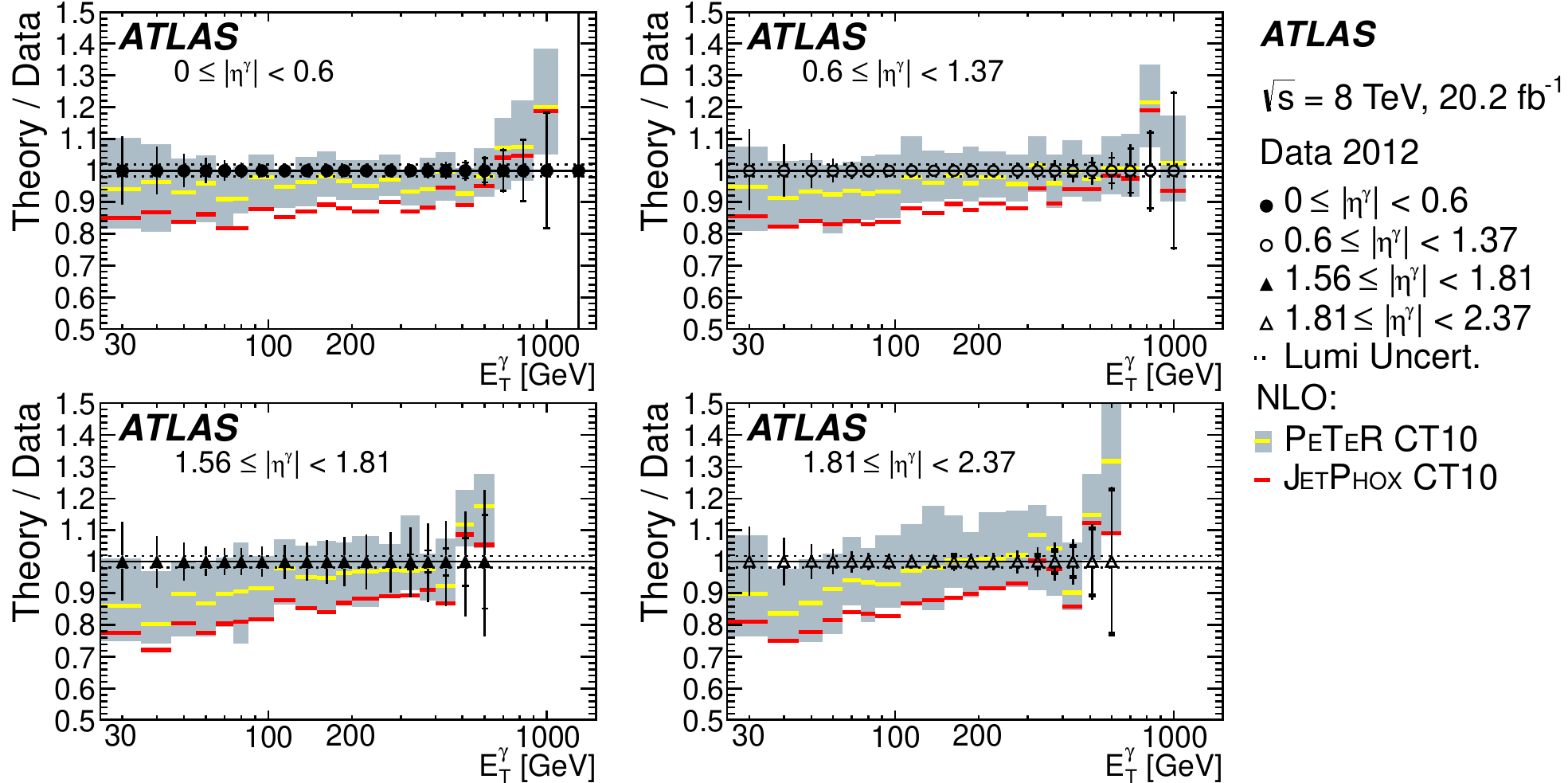}
  \caption{
    The differential photon cross-sections as a function of transverse energy ($E_\mathrm{T}$) and absolute pseudo-rapidity, using 8~TeV $pp$ collisions, are shown as a ratio of theory with data~\cite{Aad:2016xcr}.
    Theory predictions from \textsc{JetPhox} and \textsc{PeTeR} are considered.
  }
  \label{fig:photonNlo}
\end{figure}

\begin{figure}[hb]
  \centering
  \includegraphics[width=0.9\textwidth]{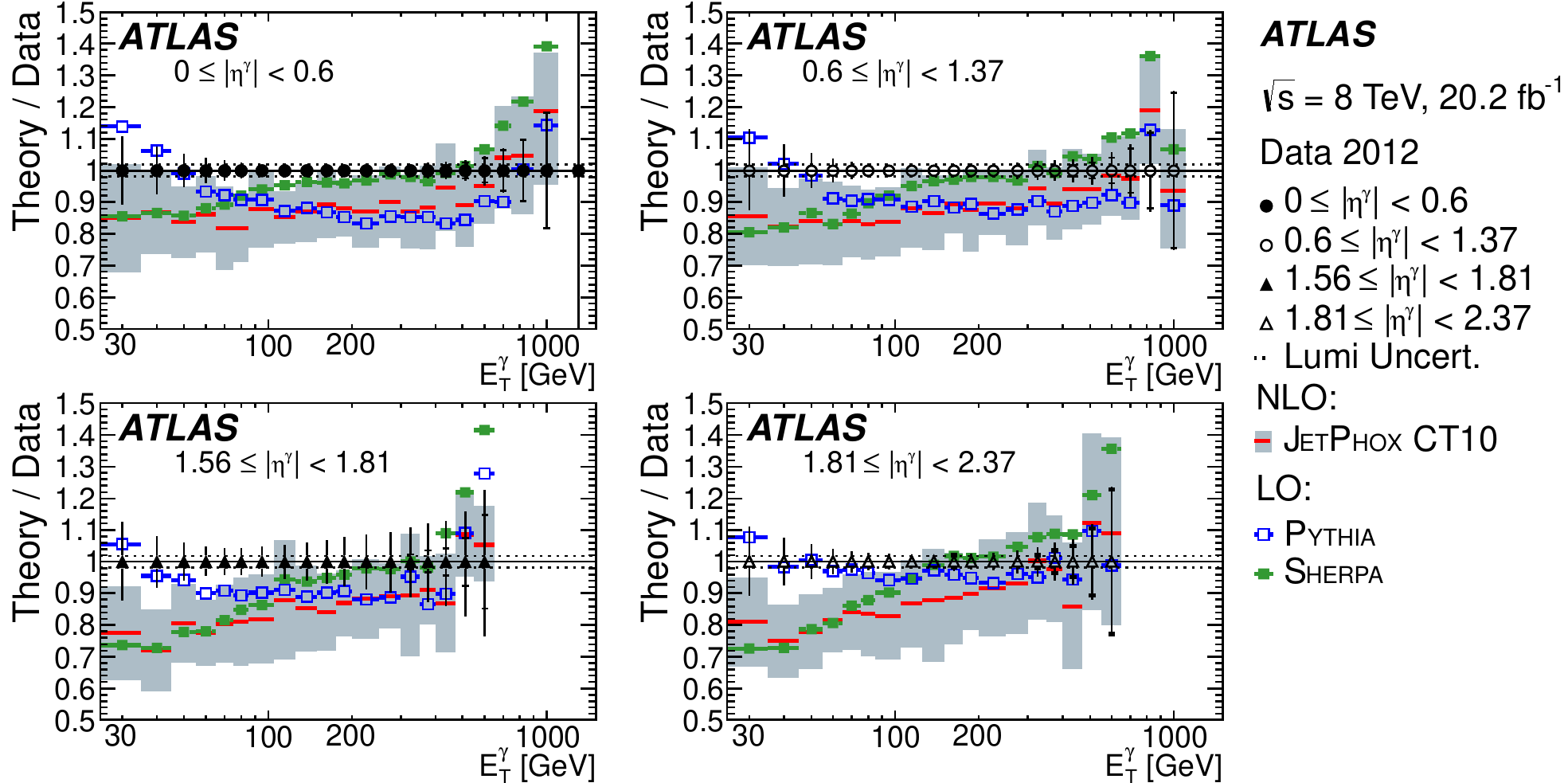}
  \caption{
    The differential photon cross-sections as a function of transverse energy ($E_\mathrm{T}$) and absolute pseudo-rapidity, using 8~TeV $pp$ collisions, are shown as a ratio of theory with data~\cite{Aad:2016xcr}.
    Theory predictions from \textsc{JetPhox}, \textsc{Pythia}, and \textsc{Herwig++} are considered.
  }
  \label{fig:photonLo}
\end{figure}

The measurement of differential cross-sections as a function of photon transverse energy and pseudo-rapidity using 8~TeV $pp$ collisions is shown in figures~\ref{fig:photonNlo} and~\ref{fig:photonLo}.
Both figures show the ratio of NLO \textsc{JetPhox}~\cite{Catani:2002ny} using the CT10 PDF set with data, while the first figure also considers the approximately NNLO prediction by \textsc{PeTeR}~\cite{Becher:2012xr}, and the second considers the LO predictions by \textsc{Pythia8} and \textsc{Herwig++}.
In general, only the \textsc{PeTeR} prediction is within 10\% of the data over the kinematic range where statistical uncertainty is not dominant.
Of the LO predictions, \textsc{Pythia} is within 10\% the data for larger values of absolute pseudo-rapidity.

\clearpage

\section{Summary}

Several recent jet and photon measurements from ATLAS are summarized, providing good tests of perturbative calculations, PDF models, and fragmentation/hadronisation simulation in MC generators.
While a good description of jet properties such as charged-particle multiplicity and charge is seen, there is no single MC generator which performs best in all situations.

As the LHC moves into a mode of operation which focuses on collecting a larger data sample, jet and photon cross-sections will become increasingly important.
The experimental uncertainties are already similar to the PDF uncertainties, and both will be reduced further when the full analysis of the 13~TeV $pp$ collision sample is complete.
This highlights the importance of the NNLO predictions that will soon be available, which will reduce the now non-negligible renormalisation and factorisation scale uncertainty.

Going forward, the properties and precision measurements of jets and photons will be crucial for fine-tuning descriptions of the Standard Model, and providing precise background estimates for new physics searches.
In summary, it is an exciting time for both jet and photon physics.

\end{document}